\documentclass[aps,pra,preprint,superscriptaddress,longbibliography]{revtex4-1}
\usepackage{amsmath,amsthm,amsfonts,amssymb,xcolor,graphicx,tikz,color,longtable,array,dsfont}
\usepackage[pdftex]{hyperref} 

\begin{document}
	
\title{Optical non-Hermitian para-Fermi oscillators}

\author{S. Rodr\'iguez-Walton}
\affiliation{Tecnologico de Monterrey, Escuela de Ingenier\'ia y Ciencias, Ave. Eugenio Garza Sada 2501, Monterrey, N.L., Mexico, 64849.}
	
\author{B. Jaramillo \'Avila}
\affiliation{CONACYT-Instituto Nacional de Astrof\'{i}sica, \'{O}ptica y Electr\'{o}nica, Calle Luis Enrique Erro No. 1. Sta. Ma. Tonantzintla, Pue. C.P. 72840, Mexico.}
	
\author{B. M. Rodr\'iguez-Lara}
\affiliation{Tecnologico de Monterrey, Escuela de Ingenier\'ia y Ciencias, Ave. Eugenio Garza Sada 2501, Monterrey, N.L., Mexico, 64849.}
\affiliation{Instituto Nacional de Astrof\'{\i}sica, \'Optica y Electr\'onica, Calle Luis Enrique Erro No. 1, Sta. Ma. Tonantzintla, Pue. CP 72840, Mexico.}

\date{\today}

\begin{abstract}
	We present a proposal for the optical simulation of para-Fermi oscillators in arrays of coupled waveguides.
	We use a representation that arises as a deformation of the $su(2)$ algebra. 
	This provides us with a set of chiral and a zero-energy-like normal modes.
	The latter is its own chiral pair and suggest the addition of controlled losses/gains following a pattern defined by parity. 
	In these non-Hermitian para-Fermi oscillators, the analog of the zero-energy mode presents the largest effective loss/gain and it is possible to tune the system to show sequences of exceptional points and varying effective losses/gains.
	These arrays can be used for mode suppression or enhancement depending on the use of loss or gain, in that order. 
	We compare our coupled mode theory predictions with finite element method simulations to good agreement. 
\end{abstract}

\maketitle

\section{Introduction}

Heisenberg quantized the harmonic oscillator using the equations of motion of its canonical variables \cite{Heisenberg1944}. 
Wigner realized that deformed oscillators can generate the same equations of motion with different underlying commutation relations \cite{Wigner1950}.
In particular, parity or reflection based deformations \cite{Yang1951} became of interest as the statistical properties arising from these represent so-called para-particles \cite{Green1953} although, in time, these were proven infeasible to detect in nature \cite{Greenberg1965}.
Recently, the quantum simulation of para-particles has fueled research in diverse avenues, such as fractional quantum Hall effect \cite{Macdonald1990,Radu2008}, Majorana particle simulation \cite{Casanova2011} and para-particle simulation \cite{Huerta2017a,Huerta2017b,Huerta2018}. 

Here, we propose an optical simulation of para-Fermi oscillators using laser-inscribed arrays of evanescently coupled waveguides \cite{Davis1996,Blomer2006,Szameit2006,Heinrich2009,Szameit2010}.
Coupled mode theory accurately describes this optical simulation platform via a coupled mode matrix whose diagonal and off-diagonal elements correspond to the localized field propagation constants and inter-waveguide coupling strengths \cite{McIntyre1973}.
We focus on para-Fermi oscillators because there exists a finite representation of these objects given in terms of a deformed $su(2)$ algebra \cite{Plyushchay1997}.
The optical simulation of the latter in laser-inscribed arrays is well understood \cite{PerezLeija2013,RodriguezLara2014,Villanueva2015} and its deformation via losses has proved useful to simulate $\mathcal{PT}$-symmetric phenomena \cite{Rodriguez2015b}.

In the following, we introduce Plyushchay representation of the para-Fermi algebra, construct a para-Fermi analog of the $\hat{J}_{x}$ oscillator, and calculate its proper energies and states.
The energies show a symmetric square root distribution and the states form chiral pairs corresponding to symmetric energies plus a single zero-energy state that is its own chiral pair. 
Next, we propose an optical simulation based on arrays of coupled waveguides and provide an explicit example for para-Fermi oscillator of order four. 
Finite element simulations show that the propagation constants and normal modes of our optical device correspond to the proper energies and states of the quantum model.
This correspondence suggests the introduction of localized loss (gain) that leave the optical analog of the zero-energy state as normal mode of the new deformed optical device that we call a non-Hermitian para-Fermi oscillator. 
This non-Hermitian mode selective behavior is inspired by different amplification or loss schemes based on symmetric zero modes\cite{Ge2017,Pan2018,Qi2018,Rivero2019}.
The optical analog of the zero-energy state displays the highest loss (gain) and it is possible to design the optical device to produce a series of effective losses (gains) as well as exceptional points where the dimension of the device is effectively reduced by one.
This motivates us to propose our optical non-Hermitian para-Fermi oscillator as mode suppressor or enhancer for engineered loss or gain, in that order. 
We close with a brief summary and conclusion.

\section{Quantum para-Fermi oscillator}

In order to deal with finite dimensional devices, we use Plyushchay representation of the para-Fermi algebra \cite{Plyushchay1997}, 
\begin{eqnarray}\label{eq:commutation}
[ \hat{I}_{+}, \hat{I}_{-} ] = 2 \hat{I}_{0} ~ \hat{\Pi},
\qquad
[ \hat{I}_{0}, \hat{I}_{\pm} ] = \pm \hat{I}_{\pm},
\end{eqnarray}
where the operators $\hat{I}_{\pm}, \hat{I}_{0}$ play a role analogous to the angular momentum generators $\hat{J}_{\pm}, \hat{J}_{3}$, $[ \hat{J}_{+}, \hat{J}_{-} ] = 2 \hat{J}_{3}$ and $[ \hat{J}_{3}, \hat{J}_{\pm} ] = \pm \hat{J}_{\pm}$, with the addition of a parity-like operator $\hat{\Pi}$.
We choose an orthonormal basis of para-Fermi states $\vert p;m\rangle$, with $m = -p, -p+1,\ldots,p-1,p$, that diagonalize the $\hat{I}_{0}$ operator,
\begin{equation}
\begin{aligned}
 \hat{I}_{0} ~\vert p; m \rangle &= m ~\vert p; m \rangle, \\
 \hat{I}_{+} ~\vert p; m \rangle &= \phi(p,m+1) ~\vert p; m+1 \rangle, \\
 \hat{I}_{-} ~\vert p; m \rangle &= \phi(p,m) ~\vert p; m-1 \rangle, \\
 \hat{\Pi} ~\vert p; m \rangle &= (-1)^{p+m} ~\vert p; m \rangle, 
\end{aligned}
\end{equation}
with the structure function,
\begin{eqnarray}\label{eq:structure}
\phi(p,m) = \sqrt{ \left( p + \frac{1}{2} \right) + \left( m - \frac{1}{2} \right) (-1)^{p+m} }.
\end{eqnarray}
This $(2p+1)$-dimensional representation of the para-Fermi algebra, corresponding to para-Fermions of order $2p$, is central to our optical simulation. 

In particular, we are interested in an equivalent of the so-called $\hat{J}_{x}$ Hamiltonian \cite{PerezLeija2013,RodriguezLara2014,Villanueva2015},
\begin{eqnarray}\label{eq:hamiltonian}
\hat{H} = g \left( \hat{I}_{+} + \hat{I}_{-} \right), 
\end{eqnarray}
that couples neighbouring para-Fermi states with a coupling strength $g$. 
Its spectrum contains nondegenerate symmetric and a zero energies \cite{Huerta2018},
\begin{eqnarray}\label{eq:eigenvalues}
\lambda_{\pm j} = \pm 2 g \sqrt{j}
\qquad \text{ for } j=0,1,2,\ldots,p,
\end{eqnarray}
for chiral states,
\begin{eqnarray} \label{eq:eigenstates}
\vert \lambda_{\pm j} \rangle = \sum_{m=-p}^{p} c_{\pm j,m} \vert p; m \rangle, \qquad \vert \lambda_{+ j} \rangle = \hat{\Pi} \vert \lambda_{- j} \rangle.
\end{eqnarray}
These states are defined in terms of probability amplitudes fulfilling the following recurrence relation,
\begin{eqnarray}\label{eq:recurrence}
c_{\pm j, m+1} = 
\frac{
		-\phi(p,m) ~c_{\pm j, m-1} 
		+ \lambda_{\pm j}~ c_{\pm j, m} 
}{
	\phi(p,m+1)
}, 
\end{eqnarray}
where $c_{\pm j, -p-1} = 0$ and $c_{\pm j, +p+1} = 0$. 
Chirality and zero-energy states play a crucial role in our extended optical model. 
It is interesting that while the so-called $\hat{J}_{x}$ Hamiltonian shows commensurable spectrum, $\pm 2 g j$, our para-Fermi Hamiltonian presents both commensurable and incommensurable sections. 
Therefore, evolution can show regular and ergodic behavior.

\section{Optical para-Fermi oscillator}

\begin{figure}[h!]
	\centering
	\includegraphics{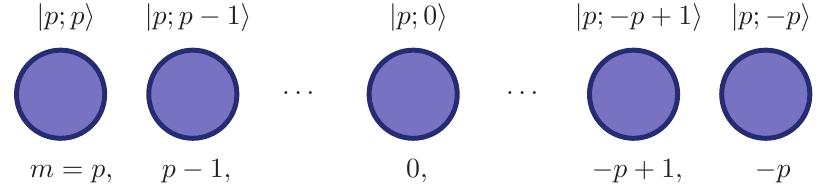}
	\caption{Schematic for the optical simulation of para-Fermi oscillators of the $\hat{I}_{x}$ type in an array of evanescently coupled single-mode waveguides.}\label{fig:1}
\end{figure}

Underlying symmetries in the form of a Lie group suggest optical simulations with arrays of coupled waveguides \cite{Rodriguez2015a,Rodriguez2018c}.
We consider an array of identical waveguides, each supporting a LP$_{01}$ mode, and use coupled mode theory to describe the propagation of their modal amplitudes.
In a manner similar to the $\hat{J}_{x}$ array \cite{PerezLeija2013,RodriguezLara2014,Villanueva2015}, the complex field amplitude at each waveguide corresponds to the complex probability amplitude for a given para-Fermi state. 
The leftmost (rightmost) waveguide corresponds to the para-Fermi state $\vert p; p \rangle$ ($\vert p; -p \rangle$) and so on, Fig. \ref{fig:1}.

\begin{figure}[h!]
	\centering
	\includegraphics{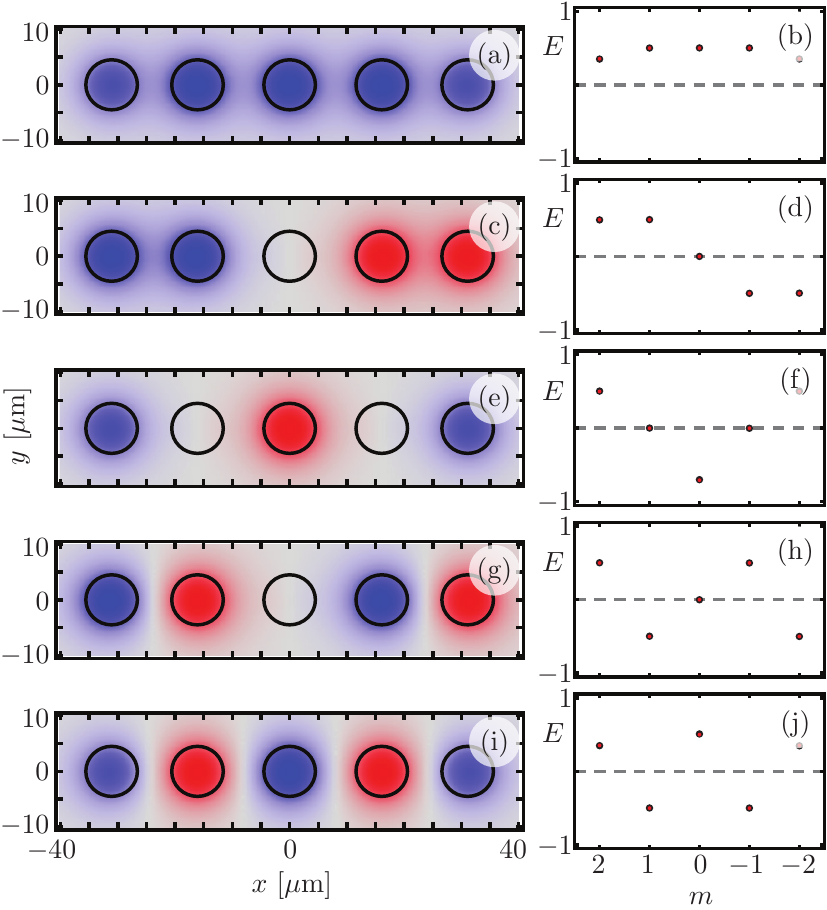}
	\caption{Normal modes for an optical para-Fermi oscillator with $p=2$. On the left, we show the spatial mode structure provided by numerical FEM. On the right, we compare our analytic CMT results for the field amplitudes, red small points, with the average field in each core from numerical FEM, black large points.}\label{fig:2}
\end{figure}

For the sake of simplicity, we explicitly deal with an example for para-Fermi oscillators of order four, or $p=2$, that yield an optical array of dimension five with coupled mode equations, 
\begin{eqnarray}
- i \frac{\mathrm{d}\vec{E}(z)}{\mathrm{d}z} = 
	g
	\underbrace{	\left(
	\begin{array}{ccccc}
		0			& 2			& 0			& 0			& 0			\\
		2			& 0			& \sqrt{2}	& 0			& 0		\\
		0			& \sqrt{2}	& 0			& \sqrt{2}	& 0		\\
		0			& 0			& \sqrt{2}	& 0			& 2		\\
		0			& 0			& 0			& 2			& 0
	\end{array}
	\right)
}_{\mathbf{M}} 
\cdot 
\underbrace{
	\left(
	\begin{array}{c}
		E_{2}(z) \\ E_{1}(z) \\ E_{0}(z) \\ E_{-1}(z) \\ E_{-2}(z)
	\end{array}
	\right)
}_{\vec{E}(z)}
\end{eqnarray}
where the mode coupling matrix, $\mathbf{M}$, is analogous to the Hamiltonian, $\hat{H}$, in Eq. (\ref{eq:hamiltonian}) and the complex field amplitudes vector, $\vec{E}(z)$, is analogous to the state vector of the quantum system, $\vert \psi(t) \rangle$. 
We obviate the single mode localized field propagation constant $\beta_{0}$ as it only introduces a common phase $e^{i \beta_{0} z}$ to the field amplitudes. 
The reduced effective propagation constants of our optical para-Fermi oscillator, 
\begin{eqnarray}
\beta_{\pm j} = \pm 2 g \sqrt{j} 
\qquad \text{ for } j=0,1,2,
\end{eqnarray}
and their corresponding normal modes,
\begin{eqnarray}
\begin{aligned}
\vec{E}_{\pm2} = \frac{1}{2\sqrt{2}}
\left(\begin{array}{c}
1 \\ \pm\sqrt{2} \\ \sqrt{2} \\ \pm \sqrt{2} \\ 1
\end{array}\right),
\qquad
\vec{E}_{\pm 1} &=& \frac{1}{2}
\left(\begin{array}{c}
1 \\ \pm 1 \\ 0 \\ \mp 1 \\ -1
\end{array}\right),
\qquad
\vec{E}_{0} = \frac{1}{2}
\left(\begin{array}{c}
1 \\ 0 \\ -\sqrt{2} \\ 0 \\ 1
\end{array}\right),
\end{aligned}
\end{eqnarray}
are the optical analog of the energy spectrum, Eq. (\ref{eq:eigenvalues}), and proper states, Eq. (\ref{eq:eigenstates}).
As expected, the normal modes form chiral pairs, 
\begin{eqnarray}
\mathbf{\Pi} \cdot \vec{E}_{\pm m} = \vec{E}_{\mp m},
\end{eqnarray}
where the para-Fermi parity matrix of order four takes the diagonal form $\mathbf{\Pi} = \mathrm{diag}(1,-1,1,-1,1)$, and the optical analog to the zero-energy state is its own chiral partner, $\mathbf{\Pi} \cdot \vec{E}_{0} = \vec{E}_{0}$.
Our finite element simulations use arrays of identical circular-cylinder waveguides with core radius $r_{co}=4.5~\mu\mathrm{m}$ and core (cladding) refractive index $n_{co}=1.4479$ ($n_{cl}=1.4440$). 
We use telecomm C-band, $\lambda = 1550~\mathrm{nm}$, to approximate a common propagation constant $\beta_{0} = 5.859\,75 \times 10^{-6}~\mathrm{rad/m}$ for the localized modes at each waveguide and coupling strengths $g_{-2,-1}=g_{1,2}= 256.636~\mathrm{rad/m}$ and $g_{-1,0}=g_{0,1}= 181.469~\mathrm{rad/m}$, given by the center-to-center core separations $d_{-2,-1}=d_{1,2}= 15.0~\mu\mathrm{m}$ and $d_{-1,0}=d_{0,1}= 16.0842~\mu\mathrm{m}$, in that order.
Table \ref{tab:1} and Fig. \ref{fig:2} show a comparison between our analytic coupled mode results with finite element simulations to good agreement.

\begin{table}
	\centering
	\begin{tabular}{c|cc}
		& \multicolumn{2}{c}{{$\beta_{0} + \beta_{j}$} ~~\footnotesize{$\left[\times 10^{6}~\mathrm{rad}/\mathrm{m}\right]$}}	\\
		\hline
		\hspace{0.5em} $j$ \hspace{0.5em}
		& \hspace{1.5em} CMT \hspace{1.5em}
		& \hspace{1.5em} FEM \hspace{1.5em}	\\
		\hline		
		-2 		& $5.85939$ 	& $5.85928$		\\
		-1		& $5.85950$		& $5.85943$		\\
		 0		& $5.85975$		& $5.85975$		\\
		 1		& $5.86001$		& $5.86003$		\\
		 2		& $5.86012$		& $5.86014$		\\
	\end{tabular}
\caption{Normal mode propagation constants for an optical para-Fermi oscillator of order four, or $p=2$, provided by analytic coupled mode theory (CMT), $\beta_{0} + \beta_{j}$, and numerical finite element method (FEM).} \label{tab:1}
\end{table}

\section{Optical non-Hermitian extension}

\begin{figure}[h!]
	\centering
	\includegraphics{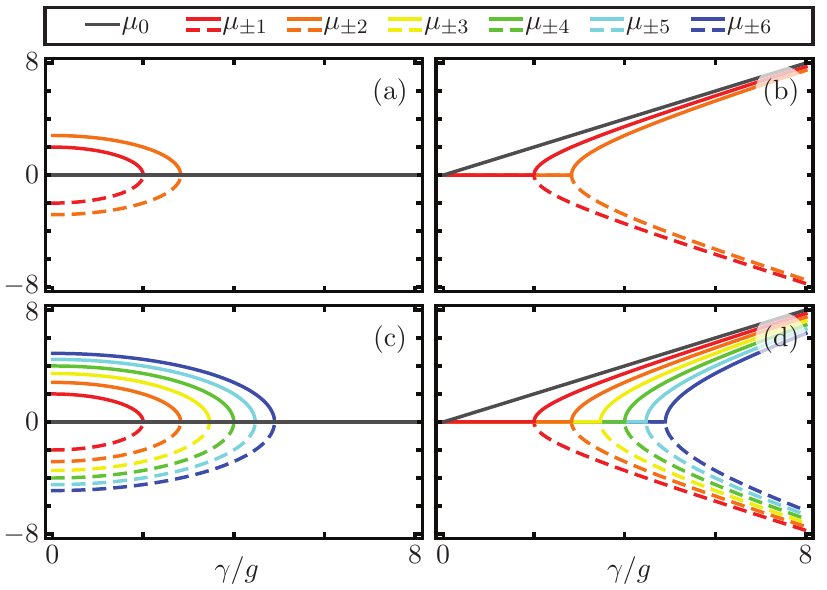}
	\caption{Real, (a), (c), and imaginary, (b), (d), parts of the scaled reduced propagation constants $\mu_{\pm j}/g$ as a function of the loss to coupling ratio $\gamma/g$ for $p=2$, (a) and (b), and $p=6$, (c) and (d).}\label{fig:3}
\end{figure}

The optical analogies to chiral pair and zero-energy states in the quantum para-Fermi oscillator inspire us to explore analogies to quantum non-Hermitian mechanics. 
As the optical analog of the zero-energy state is its own chiral pair, we introduce linear optical processes where this mode remains as a normal mode of the new coupled mode matrix,
\begin{eqnarray}\label{eq:nh}
\mathbf{M}_{nH} = \left( \beta_{0} + i \gamma_{0} \right) \mathbf{1} + i \gamma ~ \mathbf{\Pi} + g \left( \mathbf{I}_{+} + \mathbf{I}_{-} \right),
\end{eqnarray}
where the total localized loss or gain in waveguide $m$ is given by $\gamma_{0} + (1)^{m} \gamma$. 
Again, the common propagation constant, $\beta_{0}$, and a common loss/gain, $\gamma_{0}$, introduce an overall phase and an exponential loss/gain. 
In contrast, the second term can be interpreted as alternating effective losses and gains in the waveguides. 
This non-Hermitian coupled mode matrix mixes pairs of chiral partners from the optical para-Fermi oscillator,
\begin{eqnarray} \label{eq:dimmer}
\mathbf{M}_{nH} \cdot \vec{E}_{\pm j} = \left( \beta \pm \beta_{j} + i \gamma_{0} \right) \vec{E}_{\pm j} + i \gamma ~\vec{E}_{\mp j}.
\end{eqnarray}
The optical analog of the zero-energy state is its own chiral partner and, therefore, it is an normal mode of the non-Hermitian oscillator as well. 
The reduced propagation constants are
\begin{eqnarray}
\begin{aligned}
\mu_{0} &= i \gamma, \\
\mu_{\pm j} &= \pm \sqrt{ \beta_{j}^{2} - \gamma^{2} } 
\qquad 
\text{ for } j=1,\ldots,p
\end{aligned}
\end{eqnarray}
and their respective normal modes are
\begin{eqnarray}
\begin{aligned}
\vec{\mathcal{E}}_{0} &= \vec{E}_{0}, \\
\vec{\mathcal{E}}_{\pm j} &= 
\mathcal{N}_{\pm j}
\left[
	\left( \beta_{j} + \mu_{\pm j} \right) \vec{E}_{j}
	+ i \gamma ~\vec{E}_{-j} 
\right]
\end{aligned}
\end{eqnarray}
for $j=1,\ldots,p$ and where $\mathcal{N}_{\pm j}$ is a normalization constant. 
The number of reduced imaginary propagation constants depends on the interplay between coupling strength and effective loss parameter, $\mu_{\pm j} = \pm ( 4 g^2 j - \gamma^{2} )^{1/2}$. 
This is akin to symmetry breaking in a $\mathcal{PT}$ dimer. 
In our case, the dimer is formed by the chiral pairs $\{ \vec{E}_{+j}, \vec{E}_{-j} \}$ in Eq. (\ref{eq:dimmer}). 
When the reduced non-Hermitian propagation constants of a dimer are real, the normal modes $\{ \vec{\mathcal{E}}_{+j}, \vec{\mathcal{E}}_{-j} \}$  are linearly independent; when they are zero, the dimension of the dimer is reduced to one; when they are imaginary, the normal modes are, again, linearly independent.
Regardless of the relationship between the parameters of the system, the optical analog of the zero-energy mode in the Hermitian device has the reduced propagation constant with the largest (smallest) imaginary part, $\mu_{0} = i \gamma $. 
Therefore, it has larger losses (gains) than the rest of the normal modes.
Figure \ref{fig:3} displays the behaviour of the reduced propagation constants as as a function of the ratio between the loss parameter and the coupling strength, $\gamma/g$.

\begin{figure}[h!]
	\centering
	\includegraphics{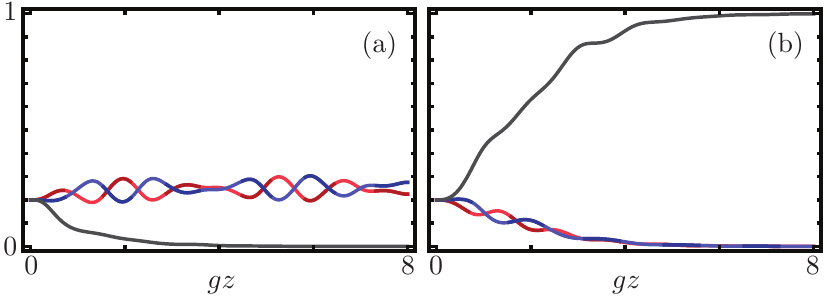}
	\caption{Propagation of the squared absolute value of the renormalized modal amplitudes, $a_{j} = \vec{\mathcal{E}_{j}}^{\dagger} \cdot \vec{E}(z) / \vert \vec{E}(z) \vert$, for effective (a) loss and (b) gain $\vert \gamma \vert = g$ for an optical non-Hermitian para-Fermi oscillator of order four, or $p=2$.}\label{fig:4}
\end{figure}

The fact that we can design our non-Hermitian para-Fermi oscillator to support anything between a single effective loss (gain) normal mode and up to $p + 1$ effective loss (gain) and $p$ gain (loss) normal modes, Fig. \ref{fig:3}, suggests its use to suppress or enhance these effective loss or gain modes, in that order. 
For example, let us consider an optical device with characteristics identical to those described before and add $2.22911~\mathrm{dB/mm}$ losses on even sites such that the common and effective loss parameters are identical $\gamma_{0} = \gamma = 256.6336~\mathrm{rad/m}$ and fulfill $0 < \gamma / g < 4$.
In this case, the optical analog of the zero-energy mode for the Hermitian device is the non-Hermitian device normal mode with the largest losses. 
Therefore, it will extinguish faster than the other normal modes.
Figure \ref{fig:4}(a) shows this phenomenon through the evolution of the absolute value for the normal modes projections of the renormalized field amplitudes for an initial field equal to the balanced superposition of all the normal modes of the device.
Figure \ref{fig:4}(b) shows the inverse process, the optical analog of the zero-energy mode is enhanced and lifted above all other modes if we introduce equivalent losses (gain) on the odd sites (even sites).

\section{Conclusion}

In summary, we propose an optical simulation of a para-Fermi oscillator analog to the so calleod $\hat{J}_{x}$ optical lattice using a finite odd-dimensional representation of the para-Fermi algebra.
These reduced effective propagation constants are distributed symmetrically on the real line and the normal modes corresponding to symmetric pairs form chiral pairs. 
There always exists an optical analog of a zero-energy mode that is its own chiral pair.
These propagation constants show both commensurable and incommensurable sectors allowing for regular and ergodic dynamics. 

We also show that, introducing loss or gains following a parity pattern determined by the position of the waveguides that leaves the optical analog of the zero-energy mode a normal mode of the deformed device, it is possible to propose an optical simulation of a non-Hermitian para-Fermi oscillator.
In this device, the optical analog of the zero-energy mode has the largest reduced effective loss or gain and the loss to coupling parameters ratio control the number of normal modes showing symmetric effective loss and gain through.
Therefore, we can use our non-Hermitian device to suppress or enhance normal modes components of impinging field distributions. 

\section*{Funding}
B.J.A. acknowledges financial support from CONACYT under C\'atedra Grupal \#551. 
B.M.R.L. acknowledges financial support from CONACYT under project CB-2015-01/255230, and the Marcos Moshinsky Foundation under 2018 Marcos Moshinsky Young Researcher Chair. 


%

\end{document}